\documentclass[conference,a4paper]{IEEEtran}

\usepackage{pgfplots}
\usepgfplotslibrary{groupplots}
\pgfplotsset{compat=newest}
\usepackage{tikz}
\usetikzlibrary{arrows,matrix,positioning,patterns,decorations,calc}
\usepackage[utf8]{inputenc}

\usepackage[innermargin=0.575in,outermargin=0.575in,top=0.545in,bottom=.55in]{geometry}

\usepackage[english]{babel}
\usepackage[T1]{fontenc}
\usepackage{epsfig}
\usepackage{amsmath, amssymb, amsbsy,bbm}
\usepackage{mathdots}
\usepackage{xspace}
\usepackage[noend]{algpseudocode}

\usepackage[linesnumbered,ruled,vlined,titlenumbered]{algorithm2e}
\usepackage{algorithmicx}
\usepackage{color}
\usepackage{cite}
\usepackage{booktabs}
\usepackage{verbatim}
\usepackage{url}
\usepackage{lipsum}
\usepackage[inline]{enumitem}
\usepackage{colortbl}
\usepackage{amsthm}
\usepackage{nicefrac}
\usepackage[keeplastbox]{flushend}
\usepackage{dblfloatfix}

\makeatletter
\newcommand\footnoteref[1]{\protected@xdef\@thefnmark{\ref{#1}}\@footnotemark}
\makeatother

\newtheorem{theorem}{Theorem}
\newtheorem{lemma}[theorem]{Lemma}
\newtheorem{corollary}[theorem]{Corollary}

\newtheorem{remark}{Remark}

\newenvironment{mymatrix}{\begin{bmatrix}} {\end{bmatrix} }

\def\ve#1{{\mathchoice{\mbox{\boldmath$\displaystyle #1$}}%
              {\mbox{\boldmath$\textstyle #1$}}%
              {\mbox{\boldmath$\scriptstyle #1$}}%
              {\mbox{\boldmath$\scriptscriptstyle #1$}}}}

\newcommand{\todo}[1]{{\color{red}[#1]}}

\definecolor{darkgreen}{rgb}{0,0.6,0}
\newcommand{\shortlongversion}[2]{#2}

\newcommand{\y}{\ve{y}}

\newcommand{\e}{\ve{e}}

\renewcommand{\e}{\ve{e}}

\newcommand{\R}{\ve{R}}

\newcommand{\F}{\mathbb{F}}

\newcommand{\ZZ}{\mathbb{Z}}

\newcommand{\x}{\ve{x}}

\newcommand{\Q}{\ve{Q}}

\renewcommand{\d}{\ve{d}}
\newcommand{\Exp}{\mathbb{E}}
\newcommand{\Var}{\mathrm{Var}}

\newcommand{\Rout}{R_\mathrm{out}}
\newcommand{\Rin}{R_\mathrm{in}}
\newcommand{\Rix}{R_\mathrm{ix}}

\newcommand{\kout}{k_\mathrm{out}}

\newcommand{\dplus}{{d^{+}}}
\newcommand{\multidp}{{B(\dplus,\d)}}
\newcommand{\dset}{\mathcal{D} }
\newcommand{\dplusset}{\mathcal{D} ^ {d}}
\newcommand{\dpevent}{\mathcal{P}_{d}}
\newcommand{\dpeventplus}{\mathcal{P}_{\dplus}}
\newcommand{\NN}{\ensuremath{\mathbb{N}}}
\newcommand{\RR}{\ensuremath{\mathbb{R}}}
\newcommand{\epspr}{\ensuremath{\epsilon}}

\newcommand{\dpmax}{d^{+}_\mathrm{max}}
\newcommand{\dplussetmax}{\mathcal{D} ^+ _\mathrm{max}}

\newcommand{\DEC}{\ensuremath{\mathsf{dec}}}

\newcommand{\dnaA}{\mathsf{A}}
\newcommand{\dnaC}{\mathsf{C}}
\newcommand{\dnaG}{\mathsf{G}}
\newcommand{\dnaT}{\mathsf{T}}

\newcommand{\SucE}{\ensuremath{\mathcal{S}}}
\newcommand{\PoiE}{\ensuremath{\mathcal{P}}}

\newcommand{\cY}{\mathcal{Y}}

\newcommand{\MC}{M_\mathrm{C}}
\newcommand{\MIn}{M_\mathrm{In}}
\newcommand{\MIx}{M_\mathrm{Ix}}

\definecolor{darkred}{rgb}{0.7,0,0}
\definecolor{indigo}{rgb}{0.29,0,0.51}

\title{Achievable Rates of Concatenated Codes\\ in DNA Storage under Substitution Errors\vspace{-.2cm}}

\author{
	\IEEEauthorblockN{\textbf{Andreas Lenz}\IEEEauthorrefmark{1},
		\textbf{Lorenz Welter}\IEEEauthorrefmark{1},
		and \textbf{Sven Puchinger}\IEEEauthorrefmark{2},}
	\IEEEauthorblockA{\IEEEauthorrefmark{1}%
		Technical University of Munich,
		andreas.lenz@mytum.de,~lorenz.welter@tum.de}
	\IEEEauthorblockA{\IEEEauthorrefmark{2}%
		Technical University of Denmark, 
		svepu@dtu.dk}\vspace{-1.0cm}
	
	\thanks{This work has been supported by the European Union's Horizon 2020 research and innovation programme under the the European Research Council (ERC) grant agreement no.~801434 and the Marie Sklodowska-Curie grant agreement no.~713683.}}

\IEEEoverridecommandlockouts

\begin{document}

\bstctlcite{IEEEexample:BSTcontrol}

\maketitle

\begin{abstract}
	In this paper, we study achievable rates of concatenated coding schemes over a deoxyribonucleic acid (DNA) storage channel. Our channel model incorporates the main features of DNA-based data storage. First, information is stored on many, short DNA strands. Second, the strands are stored in an unordered fashion inside the storage medium and each strand is replicated many times. Third, the data is accessed in an uncontrollable manner, i.e., random strands are drawn from the medium and received, possibly with errors. As one of our results, we show that there is a significant gap between the channel capacity and the achievable rate of a standard concatenated code in which one strand corresponds to an inner block. This is in fact surprising as for other channels, such as $q$-ary symmetric channels, concatenated codes are known to achieve the capacity. We further propose a modified concatenated coding scheme by combining several strands into one inner block, which allows to narrow the gap and achieve rates that are close to the capacity.
\end{abstract}


\section{Introduction}
DNA has evolved to a competitive medium for long-term archival storage. Recent experiments \cite{church_next-generation_2012,goldman_towards_2013,grass_robust_2015,yazdi_rewritable_2015,blawat_forward_2016,bornholt_dna-based_2016,erlich_dna_2017,yazdi_portable_2017,organick_random_2018,chandak_improved_2019}
have shown that it is possible to store large amounts of data using these macromolecules. These studies have addressed different important aspects of long-term data storage, such as portability \cite{yazdi_portable_2017}, random-access \cite{yazdi_rewritable_2015,organick_random_2018}, durability \cite{grass_robust_2015}, reliability and scalability \cite{organick_random_2018}. A typical DNA storage system hereby consists of the three following entities. First, a synthesis machine, which artificially creates DNA strands. Such a device allows to produce strands by chaining nucleotides (adenine [$\dnaA$], cytosine [$\dnaC$], guanine [$\dnaG$] or thymine [$\dnaT$]) to arbitrary vectors. Note that due to practical limitations, it is only possible to synthesize DNA strands of length in the order of $10^2$ and $10^3$. Second, once the strands have been created, they are stored in an adequate environment and replicated many times. Third, in order to retrieve the stored archive, the data is accessed via a sequencing machine that allows to read the stored strands. The sequencer hereby reads the strand in an uncontrollable manner, i.e., it is not directly possible to select which strand to read.

Such a storage system inherently introduces a novel channel model due to its following properties. Due to the limited strand-length induced by the synthesis machine, the data needs to be split into many strands and then encoded. Depending on the amount of data stored, the number of strands is significantly larger than the length of the strands, and typical values range up to $10^7$ for large-scale archives \cite{lenz_coding_2019}. Further, the strands are stored and received in an unordered fashion, as the sequencing machine cannot control which strand should be sequenced\footnotemark. Finally, the synthesis and sequencing, or in other words, reading and writing process of DNA molecules is prone to errors, and therefore it is necessary to develop forward-error-correcting schemes that allow to cope with the inherent disorder of the read strands and possible distortions. In view of these unique properties of DNA storage systems, it is immediate that a profound understanding of the channel and associated error-correcting codes is necessary to design cost-efficient DNA-storage systems. This becomes even more apparent as one of the most prominent bottlenecks of todays DNA storage systems is the cost associated with the synthesis and sequencing \cite{tabatabaei_dna_2019}.

The channel induced by DNA-based archival storage has been discussed from different perspectives.  Constrained codes restricting long runs of homopolymers or forcing a balanced $\dnaG\dnaC$ content have, among others, been studied in \cite{schouhamer_immink_design_2018,wang_construction_2019}. Most related to our study is the work discussing channels with unordered input and output strands. Such channels have been discussed from a worst-case point of view in \cite{lenz_coding_2019,song_sequence-subset_2019,sima_coding_2019}. On the other hand, probabilistic channels have been discussed in \cite{heckel_fundamental_2017,shomorony_capacity_2019} for the case where there are no errors inside the strands \cite{heckel_characterization_2019} and for the case, where each strand is drawn exactly once \cite{shomorony_capacity_2019}. A lower bound on the capacity of the channel under discussion was derived in \cite{lenz_upper_2019} and its achievability has been proven using a typicality-like decoder together with a random coding argument in \cite{lenz_achieving_2020}. However, until now, it is unclear how to construct efficiently encodable and decodable capacity-achieving codes for this channel.

In this work, we approach the question of how to design efficiently encodable and decodable codes for the DNA-storage channel model discussed in \cite{heckel_fundamental_2017,lenz_upper_2019,lenz_achieving_2020}.
We focus on concatenated codes in this paper as they inherently match the channel model and have been used in several experiments \cite{grass_robust_2015,erlich_dna_2017,chandak_improved_2019}.
Furthermore, concatenated coding schemes achieve the capacity of other channels, in particular the $q$-ary symmetric channel \cite{forney1966concatenated}. We show, however, that due to the nature of the discussed DNA storage channel, standard concatenated codes with inner block corresponding to one DNA strand are not able to achieve the capacity with a standard hard-decision outer decoder, independent of the selected inner and outer codes.
The main result of the paper is that concatenated coding schemes \emph{can} achieve rates close to the capacity if several DNA strands constitute an inner block.
This extends the work in \cite{lenz_achieving_2020} to efficiently encodable and decodable coding schemes.



\footnotetext{{It has been shown in \cite{yazdi_rewritable_2015} that it is possible to select specific strands for sequencing by appending carefully designed primers to the DNA molecules. Here we restrict our attention to strands with the same or similar primers that do not allow for random selection.}}

%

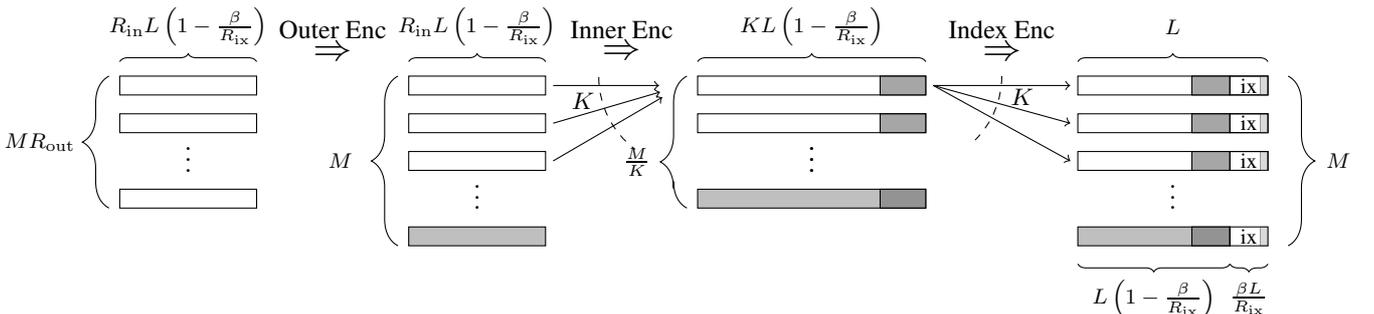
\begin{figure*}[b]
\begin{center}
\vspace{-1ex}
\begin{tikzpicture}
\def\infowidth{1.8}
\def\infoheight{0.25}
\def\outcwidth{1.8}
\def\outcheight{0.25}
\def\incwidth{3}
\def\incheight{0.25}
\def\indexwidth{2.5}
\def\indexheight{0.25}
\def\ydist{0.5}
\def\blockdistone{1.0}
\def\blockdisttwo{1.0}
\def\blockdistthree{1.0}
\coordinate (zero) at (0,0);
\coordinate (outer) at (2*\blockdistone+\infowidth,0);
\coordinate (inner) at (2*\blockdistone+\infowidth+\outcwidth+2*\blockdisttwo,0);
\coordinate (index) at (2*\blockdistone+\infowidth+\outcwidth+2*\blockdisttwo+\incwidth+2*\blockdistthree,0);

\draw [decorate,decoration={brace,amplitude=10pt,mirror,raise=4pt},yshift=0pt, color = black]
(0,0) -- (0,-3.5*\ydist) node [black,midway,xshift=-1.05cm, color = black] {\footnotesize
	$M \Rout$};
\draw [decorate,decoration={brace,amplitude=3pt,mirror,raise=4pt},yshift=0pt, color = black]
($(zero) + (1*\infowidth,0.1*\ydist)$ ) -- ($(zero)+(0,0.1*\ydist)$ ) node [black,midway,yshift=0.6cm, color = black] {\footnotesize
	$\Rin L \left(1-\frac{\beta}{\Rix}\right)$};

\draw (0,0) rectangle (\infowidth,-\infoheight);
\draw (0,-\ydist) rectangle (\infowidth,-\ydist-\infoheight);
\node at (0.5*\infowidth,-2*\ydist) {$\vdots$};
\draw (0,-3*\ydist) rectangle (\infowidth,-3*\ydist-\infoheight);
\node at (\infowidth+\blockdistone,\ydist) {\Large$\overset{\text{Outer Enc}}{\Rightarrow}$};

\draw [decorate,decoration={brace,amplitude=10pt,mirror,raise=4pt},yshift=0pt, color = black]
( $(outer) + (0,0)$) -- ( $(outer) + (0,-4.5*\ydist)$) node [black,midway,xshift=-0.9cm, color = black] {\footnotesize
	$M$};
\draw [decorate,decoration={brace,amplitude=3pt,mirror,raise=4pt},yshift=0pt, color = black]
($(outer) + (1*\outcwidth,0.1*\ydist)$ ) -- ($(outer)+(0,0.1*\ydist)$ ) node [black,midway,yshift=0.6cm, color = black] {\footnotesize
	$\Rin L \left(1-\frac{\beta}{\Rix}\right)$};

\draw ( $(outer) + (0,0)$ ) rectangle ($(outer) + (\outcwidth,-\outcheight)$);
\draw ( $(outer) + (0,-\ydist)$) rectangle ($(outer) + (\outcwidth,-\ydist-\outcheight)$);
\draw ( $(outer) + (0,-2*\ydist)$) rectangle ($(outer) + (\outcwidth,-2*\ydist-\outcheight)$);

\node at ($(outer) + (0.5*\outcwidth,-3*\ydist)$) {$\vdots$};
\draw [fill=gray,fill opacity=0.5] ($(outer)+(0,-4*\ydist)$) rectangle ($(outer)+(\outcwidth,-4*\ydist-\outcheight)$);
\node at ($(outer) + (\outcwidth + \blockdisttwo,\ydist)$) {\Large$\overset{\text{Inner Enc}}{\Rightarrow}$};

\coordinate (outerend) at ($(outer) + (\outcwidth+0.1*\blockdisttwo,0)$);
\coordinate (innerarr) at ($(inner) + (-0.2*\blockdisttwo,-0.5*\incheight)$);

\node [circle,inner sep=0pt,minimum size=6mm] (nodeinnarr) at (innerarr) {};
\draw [->] ($(outerend) + (0,-0.5*\incheight)$) -- (nodeinnarr);
\draw [->] ($(outerend) + (0,-\ydist-0.5*\incheight)$) -- (nodeinnarr);
\draw [->] ($(outerend) + (0,-2*\ydist-0.5*\incheight)$) -- (nodeinnarr);
\draw [dashed] ($(innerarr)+(0,0.45*\incheight)$) ++(0:-1.1*\blockdisttwo) arc (180:233:1.2) node[midway,pos=0.285,xshift=-0.25cm]{\small $K$};

\draw [decorate,decoration={brace,amplitude=10pt,mirror,raise=4pt,aspect=0.65},yshift=0pt, color = black]
( $(inner) + (0,0)$) -- ( $(inner) + (0,-3.5*\ydist)$) node [black,midway,xshift=-0.8cm,pos=0.65, color = black] {\footnotesize
	$\frac{M}{K}$};
\draw [decorate,decoration={brace,amplitude=3pt,mirror,raise=4pt},yshift=0pt, color = black]
($(inner) + (1*\incwidth,0.1*\ydist)$ ) -- ($(inner)+(0,0.1*\ydist)$ ) node [black,midway,yshift=0.6cm, color = black] {\footnotesize
	$KL \left(1-\frac{\beta}{\Rix}\right)$};

\draw ( $(inner) + (0,0)$ ) rectangle ($(inner) + (\incwidth,-\incheight)$);
\draw ( $(inner) + (0,-\ydist)$) rectangle ($(inner) + (\incwidth,-\ydist-\incheight)$);
\node at ($(inner) + (0.5*\incwidth,-2*\ydist)$) {$\vdots$};
\draw [fill=gray,fill opacity=0.5] ($(inner)+(0,-3*\ydist)$) rectangle ($(inner)+(\incwidth,-3*\ydist-\incheight)$);

\draw [fill=gray,fill opacity=0.7] ( $(inner) + (0.8*\incwidth,0)$ ) rectangle ($(inner) + (\incwidth,-\incheight)$);
\draw [fill=gray,fill opacity=0.7] ( $(inner) + (0.8*\incwidth,-\ydist)$) rectangle ($(inner) + (\incwidth,-\ydist-\incheight)$);
\node at ($(inner) + (0.5*\incwidth,-2*\ydist)$) {$\vdots$};
\draw [fill=gray,fill opacity=0.7] ( $(inner) + (0.8*\incwidth,-3*\ydist)$) rectangle ($(inner)+(\incwidth,-3*\ydist-\incheight)$);

\node at ($(inner) + (\incwidth + \blockdisttwo,\ydist)$) {\Large$\overset{\text{Index Enc}}{\Rightarrow}$};

\draw [->] ($(inner) + (\incwidth+0.1*\blockdistthree,-0.5*\incheight)$) -- ($(index) + (-0.1*\blockdistthree,-0.5*\indexheight)$);
\draw [->] ($(inner) + (\incwidth+0.1*\blockdistthree,-0.5*\incheight)$) -- ($(index) + (-0.1*\blockdistthree,-\ydist-0.5*\indexheight)$);
\draw [->] ($(inner) + (\incwidth+0.1*\blockdistthree,-0.5*\incheight)$) -- ($(index) + (-0.1*\blockdistthree,-2*\ydist-0.5*\indexheight)$);
\draw [dashed]($(inner)+(\incwidth+0.1*\blockdistthree,0)$) ++(0:0.9*\blockdistthree) arc (0:-45:1.2) node[midway,pos=0.3,xshift=0.3cm]{\small $K$};

\draw [decorate,decoration={brace,amplitude=10pt,mirror,raise=4pt},yshift=0pt, color = black]
( $(index) + (1.05*\indexwidth,-4.5*\ydist)$) -- ( $(index) + (1.05*\indexwidth,0)$) node [black,midway,xshift=0.8cm, color = black] {\footnotesize
	$M$};

\draw [decorate,decoration={brace,amplitude=3pt,mirror,raise=4pt},yshift=0pt, color = black]
( $(index)+(0,-4.7*\ydist)$) -- ( $(index) + (0.8*\indexwidth,-4.7*\ydist)$) node [black,midway,yshift=-0.6cm, color = black] {\footnotesize
	$L \left( 1 - \frac{\beta}{\Rix} \right)$};

\draw [decorate,decoration={brace,amplitude=3pt,mirror,raise=4pt},yshift=0pt, color = black]
( $(index)+(0.8*\indexwidth,-4.7*\ydist)$) -- ( $(index) + (1*\indexwidth,-4.7*\ydist)$) node [black,midway,yshift=-0.6cm, color = black] {\footnotesize
	$\frac{\beta L}{\Rix}$};

\draw [decorate,decoration={brace,amplitude=3pt,mirror,raise=4pt},yshift=0pt, color = black]
($(index) + (1*\indexwidth,0.1*\ydist)$ ) -- ($(index)+(0,0.1*\ydist)$ ) node [black,midway,yshift=0.6cm, color = black] {\footnotesize
	$L$};

\draw ( $(index) + (0,0)$ ) rectangle ($(index) + (\indexwidth,-\indexheight)$);
\draw [fill=gray,fill opacity=0.7] ( $(index) + (0.6*\indexwidth,0)$ ) rectangle ($(index) + (0.8*\indexwidth,-\indexheight)$);
\draw ( $(index) + (0.8*\indexwidth,0)$ ) rectangle ($(index) + (\indexwidth,-\indexheight)$) node[pos=.5] {\footnotesize ix};
\draw [fill=gray, opacity=0.3] ( $(index) + (0.96*\indexwidth,0)$ ) rectangle ($(index) + (\indexwidth,-\indexheight)$);

\draw ( $(index) + (0,-\ydist)$) rectangle ($(index) + (\indexwidth,-\ydist-\indexheight)$);
\draw [fill=gray,fill opacity=0.7] ( $(index) + (0.6*\indexwidth,-\ydist)$ ) rectangle ($(index) + (0.8*\indexwidth,-\ydist-\indexheight)$);
\draw ( $(index) + (0.8*\indexwidth,-\ydist)$ ) rectangle ($(index) + (\indexwidth,-\ydist-\indexheight)$) node[pos=.5] {\footnotesize ix};
\draw [fill=gray, opacity=0.3] ( $(index) + (0.96*\indexwidth,-\ydist)$ ) rectangle ($(index) + (\indexwidth,-\ydist-\indexheight)$);

\draw ( $(index) + (0,-2*\ydist)$) rectangle ($(index) + (\indexwidth,-2*\ydist-\indexheight)$);
\draw [fill=gray,fill opacity=0.7] ( $(index) + (0.6*\indexwidth,-2*\ydist)$ ) rectangle ($(index) + (0.8*\indexwidth,-2*\ydist-\indexheight)$);
\draw ( $(index) + (0.8*\indexwidth,-2*\ydist)$ ) rectangle ($(index) + (\indexwidth,-2*\ydist-\indexheight)$) node[pos=.5] {\footnotesize ix};
\draw [fill=gray, opacity=0.3] ( $(index) + (0.96*\indexwidth,-2*\ydist)$ ) rectangle ($(index) + (\indexwidth,-2*\ydist-\indexheight)$);

\node at ($(index) + (0.5*\indexwidth,-3*\ydist)$) {$\vdots$};

\draw [fill=gray,fill opacity=0.5] ($(index)+(0,-4*\ydist)$) rectangle ($(index)+(0.8*\indexwidth,-4*\ydist-\indexheight)$);
\draw [fill=gray,fill opacity=0.7] ( $(index) + (0.6*\indexwidth,-4*\ydist)$ ) rectangle ($(index) + (0.8*\indexwidth,-4*\ydist-\indexheight)$);
\draw ( $(index) + (0.8*\indexwidth,-4*\ydist)$ ) rectangle ($(index) + (\indexwidth,-4*\ydist-\indexheight)$) node[pos=.5] {\footnotesize ix};
\draw [fill=gray, opacity=0.3] ( $(index) + (0.96*\indexwidth,-4*\ydist)$ ) rectangle ($(index) + (\indexwidth,-4*\ydist-\indexheight)$);

\end{tikzpicture}
\end{center}
\vspace{-4ex}
\caption{Illustration of the Encoder. $\kout$ strands are encoded via a MDS outer code with rate $\Rout$ by representing them each as symbols over a large alphabet. Then a block of $K$ strands are encoded via an inner encoder by a code with rate $\Rin$ described in Lemma~\ref{lem:inner_capacity_achieving_code}. Each inner codeblock is then split again into $K$ strands. Each strand gets an index appended which is protected via an index code with rate $\Rix$.}
\label{fig:encoder}
\end{figure*}

\section{Preliminaries}\label{sec:preliminaries}
For any integer $n \in \mathbb{N}$, we write $[n] = \{1,2,\dots,n\}$ as the set of integers up to $n$. Throughout the paper, vectors are highlighted by bold letters, such as $\x = (x_1,\dots,x_n)$. 
Concerning asymptotic statements, whenever we refer to a statement to be true \emph{for large enough $n$}, we mean that there exists some $n_0$ such that the statement is fulfilled for all $n\geq n_0$. For a summary of parameters and relevant random variables, we refer the reader to Tables \ref{tab:parameters} and \ref{tab:random_variables}. For two vectors $\x,\y \in \ZZ_2^n$, we denote $d(\x,\y)$ as the Hamming distance of $\x$ and $\y$.
\subsection{Channel Model}\label{ssec:channel_model}
We start by introducing the channel model as in \cite{heckel_fundamental_2017,lenz_upper_2019,lenz_achieving_2020}. The input of the channel are $M$ strands $\x_i \in \ZZ_2^L$, $i \in [M]$, each of length $L$. Note that for simplicity we assume binary strands here, however our results can be directly extended to $q$-ary symmetric channels. From these strands, $N$ strands are drawn uniformly at randomly with replacement and perturbed by errors, resulting in output strands $\y_{j}$, $j \in [N]$ with 
$$ \y_j = \x_{I_j} + \e_j, $$
where $\e_j \sim \mathsf{Ber}(p)^L$ are random error vectors and $I_j \in [M]$ are i.i.d. uniform random draws. For our asymptotic statements, we assume $M = 2^{\beta L}$ and $N = cM$ for some fixed $0<\beta<1$ and $c \in \mathbb{R}$ and then let $M \to \infty$. This choice is due to the fact that this is practically and theoretically the most interesting and relevant region, c.f. \cite{heckel_fundamental_2017}. 
\begin{remark}
	In this model we do not incorporate deletion or insertion errors for two reasons. First, the capacity of channels with insertions or deletions is unknown and therefore also the capacity of the overall channel is unknown. Second, our main focus in this work is to analyze the effect of the disorder of the output strands. However, we believe qualitatively similar results hold also for channels with insertions and deletions. 
\end{remark}

\subsection{Channel Capacity and Related Quantities}

The capacity of the channel is given by \cite{lenz_upper_2019,lenz_achieving_2020}
$$ C = \sum_{d=0}^{\infty} p_c(d) C_d - \beta(1-\mathrm{e}^{-c}), $$
where $p_c(d) = e^{-c}\nicefrac{c^d}{d!}$ is the probability mass function (pmf) of the Poisson distribution and $C_d$ is the capacity of the $d$-multi-draw channel
\begin{align}
C_d = 1 + \sum_{i=0}^{d} B_{d,p}(i)\log_2\left( \frac{B_{d,p}(i)}{B_{d,p}(i)+B_{d,p}(d-i)}\right), \label{eq:multi-draw_capacity}
\end{align}
c.f. \cite{mitzenmacher_theory_2006}, where
$B_{d,p}(i) = \binom{d}{i} p^i (1-p)^{d-i}$ is the pmf of the binomial distribution. Hereby, the $d$-multi-draw channel is a discrete memoryless channel $\ZZ_2 \to \ZZ_2^d$, where a single input bit is received over $d$ independent binary symmetric channels with crossover probability $p$, see \cite{mitzenmacher_theory_2006,lenz_upper_2019}.
\subsection{Concatenated Coding Scheme}\label{ssec:coding_scheme}
%
We present our concatenated coding scheme that is discussed in this paper. For a visualization, see Fig.~\ref{fig:encoder}. Assume, we want to encode a data archive of $B$ bits. First, the data is split into $\kout$ strands, each of length $B/\kout$. The $\kout$ strands are then encoded via a maximum distance separable (MDS) outer code $[M,\kout]$ over $\F_{2^{\nicefrac{B}{\kout}}}$ of length $M$, dimension $\kout$ and rate $\Rout=\nicefrac{\kout}{M}$. Now, every $K$ consecutive strands are combined to one block each and encoded via an $[KB/(\kout \Rin
),KB/\kout]$ inner code to be described in detail in Lemma \ref{lem:inner_capacity_achieving_code}. Abbreviating $\ell = B/(\kout\Rin)$, each inner codeblock is again split into $K$ strands, each of length $\ell$. Then, each strand is appended with an index of $\log M = \beta L$ bits and individually protected via a multi-draw capacity-achieving index code of rate $\Rix$. Since the total strand length is $L$, we have $\ell = L(1-\nicefrac{\beta}{\Rix})$ and the overall rate is $R = \Rout\Rin(1-\nicefrac{\beta}{\Rix})$. Note that for $K=1$ we obtain standard concatenated codes, where each strand is an inner codeword.

\subsection{Decoding Procedure} \label{ssec:decoding}
The decoder acts in four stages: \begin{enumerate*} 
	\item clustering \item index decoding \item inner decoding \item outer decoding \end{enumerate*}. First, the output strands $\y_1,\dots,\y_N$ are grouped to maximal clusters $\cY_1, \dots,\cY_{\widehat{M}}$, such that for all $\y,\y'\in \cY_i$, $d(\y,\y') \leq \rho L$, where $\rho$ is the clustering diameter. This can, for example, be achieved by a greedy algorithm that selects an arbitrary strand $\cY_1 = \{\y_j\}$ and then successively adds strands to $\cY_1$ which have distance at most $\rho L$ from all sequences in $\cY_1$. This procedure goes on until there are no more output sequences that can be put into the cluster. The algorithm then continues by creating a new cluster $\cY_2$ from the remaining strands and in the same fashion. This procedure continues until each strand $\y_1,\dots,\y_N$ is assigned exactly one cluster.
%
%
%
%
We will choose $\rho \approx 2p$, as the expected number of errors from the channel are $pL$ and thus two strands that stem from the same origin have an expected distance of at most $2pL$.

Each cluster is now decoded with the index code if $C_{|\cY_i|}> \Rix$ and discarded otherwise. Then, each cluster is assigned to its position $i$ of the decoded index. If there are two or more clusters which have the same index, then discard both. In the following, let $\widehat{d}_i$ be the size of the cluster with decoded index $i$. Further abbreviate $\widehat{\d}^{(b)} = (\widehat{d}_{bK+1},\dots,\widehat{d}_{bK+K}) \in \NN_0^K$ as the draws of one block $b \in [\frac{{M}}{K}]$. Note that $\widehat{d}_{i} \neq d_{i}$ is possible due to clustering- or index-decoding errors. Now, given the decoded indices, the $K$ strands of an inner block $b$ are results of transmission over $K$ parallel multi-draw channels with draws $\widehat{\d}^{(b)}$ and the capacity of these channels is given by $C_{\widehat{\d}^{(b)}} = \nicefrac{1}{K} \sum_{j=1}^{K}C_{\widehat{d}_{bK+j}}$, as we will see in Lemma \ref{lem:inner_capacity_achieving_code}.\footnote{Note that the exact drawing realization $\widehat{\d}^{(b)}$ is random and thus unknown to the encoder for a specific inner block. This implies that by choosing a fixed inner rate, the rate is too large for weak channels and too small for stronger channels with many draws. As we will show later in Lemma \ref{lem:relative_number_of_draws_Poisson}, however, for large~$K$, the drawing distribution (and thus also the sum capacity) inside the blocks converges for a large fraction of blocks, which allows to choose the inner rate accordingly. } The inner code therefore decodes an inner block, if $C_{\widehat{\d}^{(b)}} > \Rin$ and declares an erasure otherwise as it is only possible to decode reliably if the sum capacity is smaller than the inner rate, see Lemma \ref{lem:inner_capacity_achieving_code}. Finally, the decoded inner codewords are fed into a unique decoding algorithm of the outer MDS code.

We associate the following random variables with the decoding process that will be helpful for analyzing the decoding error probability. First, $\MC$ counts the number of wrong clusters $\MC = |\{i: \nexists i' \in [M] \text{ s.t. } \cY_i = \{j:I_j=i' \}|$, i.e., the number of clusters that do not consist of exactly all output strands that stem from one input strand. Next, $\MIx$ counts the number of clusters with $C_{d_i}>\Rix$ that are decoded wrongly, assuming that all strands were clustered correctly. Similarly, $\MIn$ counts the number of inner codewords that are decoded wrongly, assuming correct clustering and index decoding. We further write $Q_\d = \{i : \d^{(i)} = \d\}$ and $\Q = (Q_\d)_{\d \in \NN_0^K}$.
For the readers convenience, we summarize all relevant parameters and random variables in Tables \ref{tab:parameters} and \ref{tab:random_variables}.
%
%

\begin{table} 
	\setlength{\tabcolsep}{4.7pt}
	\centering
	\caption{Summary of relevant parameters}
	{\renewcommand{\arraystretch}{1.1}
		\begin{tabular}{ccc} \specialrule{.8pt}{0pt}{0pt}
			Param. & Description & Relations \\ \specialrule{.8pt}{0pt}{0pt}
			$c$ & Reading rate/Reading rate & $c = \nicefrac{N}{M}$ \\
			$\beta$ & Strand density & $\beta = \nicefrac{\log M}{L}$ \\
			$p$ & Error probability of the channel & \\
			$\Rout$ & Rate of the outer code & $\Rout = \nicefrac{\kout}{M}$ \\
			$K$ & Block size of the inner code & \\
			$\Rin$ & Rate of the inner code & $R = \Rout\Rin(1-\nicefrac{\beta}{\Rix})$ \\
			$\Rix$ & Rate of the index code & \\

			 \specialrule{.8pt}{0pt}{0pt}
	\end{tabular}\label{tab:parameters}}
\vspace{-.4cm}
\end{table}
\begin{table} 
	\setlength{\tabcolsep}{1.8pt}
	\centering
	\caption{Summary of important random variables}
	{\renewcommand{\arraystretch}{1.1}
		\begin{tabular}{ccc} \specialrule{.8pt}{0pt}{0pt}
			RV & Description & Relations \\ \specialrule{.8pt}{0pt}{0pt}
			$I_j$ & Index of the $j$-th drawn strand & $\y_j = \x_{I_j} + \e_j$  \\
			$d_i$ & Num. draws of $i$-th input strand& $d_i = |\{j:I_j=i\}|$ \\
			$\d^{(i)}$ & Num. draws of strands in block $i$ & $\!\d^{(i)} \!=\! (d_{iK\!+\!1},\dots,d_{iK\!+\!K})$ \\
			$\cY_i$ & Estimated cluster of strand $i$ & $i\in[\widehat{M}]$  \\
			$\widehat{d}_i$ & Estimated num. draws/cluster sizes &  $\hat{d}_i = |\cY_i|$ \\
			$Q_\d$ & Num. blocks that are drawn $\d$ times & $Q_\d = \{i : \d^{(i)} = \d\}$ \\
			$\MC$ & Number of wrong clusters & \\
			$\MIx$ & Number of wrongly decoded indices &  \\
			$\MIn$ & Num. wrongly decoded inner codewords & \\
			 \specialrule{.8pt}{0pt}{0pt}
	\end{tabular}\label{tab:random_variables}}
\vspace{-.5cm}
\end{table}


\section{Achievable Rates Using Hard-decision Decoded Concatenated Codes}
In this section, we derive the main result of this paper: achievable rates for concatenated codes in the DNA storage channel with groups of $K$ strands as inner blocks. In Section~\ref{ssec:main_statement}, we first formulate the main statement (Theorem~\ref{thm:achievability}) and analyze the resulting achievable rate for $K \to \infty$.
The remaining sections contain the proof of Theorem~\ref{thm:achievability}: first we show three lemmas in Section~\ref{ssec:lemmas} and then conclude the proof in Section~\ref{ssec:proof_of_main_statement}.

\subsection{Main Statement and Discussion}\label{ssec:main_statement}

\begin{theorem}\label{thm:achievability}
Consider a DNA storage channel with parameters $p$, $c$, and $\beta$ as in Section~\ref{ssec:channel_model}.
Let $K \in \NN$ and $\Rix,\Rin \in (0,1)$ with $\beta<{\Rin(1-\nicefrac{\beta}{\Rix})(1-h(2p))}$ be parameters of a concatenated coding scheme as in Sections~\ref{ssec:coding_scheme} and \ref{ssec:decoding}.
Then there is such a scheme that achieves rate $R = \Rin \Rout (1-\nicefrac{\beta}{\Rix})$, for any outer rate $\Rout$ that satisfies
\begin{equation}
\Rout < \sum _{{\d \in \NN_0^K: \,C_\d(\Rix)>\Rin}} \hspace{-.5cm}  p_c(\d), \label{eq:achievable_rate}
\end{equation}
where
$p_c(\d) = \prod_{i=1}^{K} p_c(d_i)$, $p_c(d_i) = \tfrac{c^{d_i}}{d_i!} e^{-c}$
is the pmf of the $K$-variate Poisson distribution and (using the multi-draw channel capacity $C_d$, cf.~\eqref{eq:multi-draw_capacity})
\begin{align*}
C_{\d}(\Rix) = \tfrac{1}{K} \sum_{i=1}^{K} C_{d_i}(\Rix), \quad
{C}_{d} (\Rix) = \begin{cases}
C_d, &C_d>\Rix, \\
0, &\text{else.}
\end{cases}
\end{align*}
\end{theorem}

%

We prove Theorem~\ref{thm:achievability} in the next subsections, but first we discuss and interpret the result.

\begin{remark}
Using Theorem~\ref{thm:achievability}, a lower bound on the maximal achievable rate of a concatenated coding scheme with $K$ blocks can be computed by finding the supremum of the right-hand side of \eqref{eq:achievable_rate} over $\Rix$ and $\Rin$. Since $C_{d}(\Rix)$ attains only discrete values, it is only necessary to consider $\Rix = (1-\epsilon)C_{d}$ for $d=1,2,3,\dots$ and sufficiently small $\epsilon$.

It is not apparent from Theorem~\ref{thm:achievability} how to compute the sum on the right-hand side of \eqref{eq:achievable_rate} efficiently.
We will return to this issue in Section~\ref{sec:numerical_results}, where we show that it can be efficiently estimated using a simple Monte-Carlo simulation.
\end{remark}

\begin{remark}
It is in fact not possible to achieve higher rates than those given in Theorem \ref{thm:achievability} with our hard-decision decoded coding scheme. This is because choosing an outer rate violating \eqref{eq:achievable_rate}, there will, with high probability, be more erasures than the redundancy of the outer code, which leads to a decoding error.
\end{remark}

For $K \to \infty$, the achievable rate in Theorem~\ref{thm:achievability} converges to the following value.
%
\renewcommand{\thetheorem}{\arabic{theorem}.A}%
\begin{corollary}\label{cor:achievable_rate_large_k}
	Consider the setting of Theorem~\ref{thm:achievability}. For large enough $K$, any rate $R$ is achievable if it fulfills
	\begin{equation}
	R < \left(\sum_{d=0}^{\infty} p_c(d) C_d(\Rix)\right) (1-\nicefrac{\beta}{\Rix}). \label{eq:achievable_rate_large_k}
	\end{equation}
\end{corollary}
\begin{proof}
	We will prove the lemma by slightly rewriting the achievability result from Theorem \ref{thm:achievability} and the law of large numbers. To start with, we rewrite the bound on $\Rout$ of Theorem \ref{thm:achievability} in a probabilistic fashion. Let $\d = (d_1,\dots,d_K)$ be $K$ independent Poisson distributed variables with parameter $c$. Then, the bound from Theorem \ref{thm:achievability} becomes
	\begin{equation}
	\Rout < \Pr\left(\tfrac{1}{K} \sum_{i=1}^K C_{d_i}(\Rix)>\Rin \right). \label{eq:achievable_rate_probability_formulation}
	\end{equation}
	Note the here the probability is over the artificial random variables $d_1,\dots,d_K$. Now, by the law of large numbers, for any $\epsilon>0$ and large enough $K$,
	$$ \Pr\left(\left|\tfrac{1}{K} \sum_{i=1}^K C_{d_i}(\Rix)-\Exp(C_{d_i}(\Rix))\right|>\epsilon \right) < \epsilon.$$
	Setting $\Rin = \Exp(C_{d_i}(\Rix))-\epsilon$ this means that any $\Rout < 1-\epsilon$ is achievable for large enough $k$ and we obtain that any
	$$ R < \Exp(C_{d_i}(\Rix))(1-\tfrac{\beta}{\Rix}). $$
	is achievable, by choosing $\epsilon$ as small as desired. Identifying $\Exp(C_{d_i}(\Rix)) = \sum_{d=0}^{\infty} p_c(d) C_d(\Rix)$ proves the lemma.
\end{proof}
%
For $K \to \infty$, we obtain the achievable rate of a concatenated code by finding the supremum of the right-hand side of \eqref{eq:achievable_rate_large_k} over $\Rix$.
Since the $C_d(\Rix)$ change only for discrete values of $\Rix$, we have the following:
\begin{align*}
\textstyle R_\mathrm{max} &= \sup_{\Rix \in (0,1)} \textstyle\left(\sum_{d=0}^{\infty} p_c(d) C_d(\Rix)\right) (1-\nicefrac{\beta}{\Rix}) \\
&= \max_{d^* \in \ZZ_{> 0}} \sup_{\Rix \in (C_{d^*-1},C_{d^*})} \textstyle\left(\sum_{d=d^*}^{\infty} p_c(d) C_d\right) (1-\nicefrac{\beta}{\Rix}) \\
&= \max_{d^* \in \ZZ_{> 0}} \textstyle\left(\sum_{d=d^*}^{\infty} p_c(d) C_d\right) (1-\nicefrac{\beta}{C_{d^*}}),
\end{align*}
where the maximization over $d^*$ is indeed a maximization since $1-\nicefrac{\beta}{C_{d^*}} \to 1$ and the sum converges to $0$ for $d^* \to \infty$, hence the supremum is attained a finite $d^*$.
We compare $R_\mathrm{max}$ to the capacity of the DNA storage channel \cite{lenz_upper_2019,lenz_achieving_2020}.

\setcounter{theorem}{1}
\renewcommand{\thetheorem}{\arabic{theorem}.B}%
\begin{corollary}\label{cor:asymptotic_behavior_c_p}
Let $\textstyle C = \sum_{d=1}^{\infty} p_c(d) C_d - \beta(1-\mathrm{e}^{-c})$ be the capacity of the DNA storage channel \cite{lenz_upper_2019,lenz_achieving_2020}. With notation as above, we have
\begin{align*}
R_\mathrm{max}-C \to 0
\end{align*}
for $p \to 0$ (and fixed $c$) or $c \to \infty$ (and fixed $p$).
\end{corollary}

\begin{proof}
First consider $p \to 0$: We have
\begin{align*}
\textstyle R_\mathrm{max} \geq \sum_{d=1}^{\infty} p_c(d) C_d  - \beta \tfrac{\sum_{d=1}^{\infty} p_c(d) C_d}{C_1}
\end{align*}
Due to $C_d$ converges to $1$ for $p \to 0$, we have $C_1 \to 1$ and $\sum_{d=1}^{\infty} p_c(d) C_d \to 1-p_c(0) = 1-e^{-c}$.

For $c \to \infty$, for any sequence of $c \to \infty$, there is a sequence of $d^* \in \NN$ such that $d^* \to \infty$ (i.e., $C_{d^*} \to 1$) and $\sum_{d=d^*}^{\infty} p_c(d) C_d \to \sum_{d=1}^{\infty} p_c(d) C_d$. This proves the claim.
\end{proof}
\renewcommand{\thetheorem}{\arabic{theorem}}%
The numerical analysis in Section~\ref{sec:numerical_results} confirms a part of Corollary \label{cor:asymptotic_behavior_c_p} as for $c=10$, the maximal achievable rate for $K \to \infty$ is quite close to the actual capacity of the channel.

%
%

\subsection{Lemmas}\label{ssec:lemmas}

In the following, we present three lemmas that we will need for the proof of the main statement in the next subsection.
Recall the notions introduced in Section~\ref{ssec:decoding}, in particular the random variables $\MC$, $\MIx$, $\MIn$, $Q_\d$ (see also Table~\ref{tab:random_variables}).

In the concatenated scheme for the DNA storage channel discussed in Section~\ref{sec:preliminaries}, there is a fundamental difference to concatenated coding schemes designed for other channels (e.g., the $q$-ary symmetric channel): inner blocks are transmitted through \emph{different} channels.
In this case, an inner block corresponds to a block of $K$ strands and the channel it is transmitted through is a $K$-parallel multi-draw channel, where the draw vector $\d$ varies from block to block.
A major difficulty compared to channels with the same ``inner channel'' is thus to find an inner code that achieves the capacity of a large number of such channels--jointly. 
We address this question in the following lemma.

\begin{lemma}\label{lem:inner_capacity_achieving_code}
Let $K$, $\dpmax$ be integers and choose a rate $R$.
%
For any $\epsilon>0$, there is a code of rate $R$ and sufficiently large length for the family of $K$-parallel multi-draw channels such that
the decoding error probability is at most $\epsilon$ for any realization of the channel with draw vector $\d$ s. t. $\sum_{i=1} ^K d_i \leq \dpmax$ and
\begin{align*}
R &<  \frac{1}{K} \sum _{i=1} ^K  \mathrm{C}_{d_i}.
\end{align*}
%
%
\end{lemma}


\begin{proof}
We prove this Lemma using a random coding argument and a typical-sequence decoder \cite[Chapter 7]{cover_elements_2006}. Recall that the $K$-parallel multi-draw channel with draw vector $\d = (d_1,\dots,d_J)$ is $K$ parallel multidraw channels, each with $d_i$ draws. It therefore is a memoryless channel with input alphabet $\ZZ_2^K$ and output alphabet $\ZZ_2^{d_1}\times\dots\times\ZZ_2^{d_K}$. We denote each input symbol by $(X_1,\dots,X_K)$ and the corresponding output symbol by $\cY_1,\dots,\cY_K$. For this channel we specify the symbol-wise input distribution as \mbox{$\Pr\big((X_1,\dots,X_K) = (x_1,\dots,x_K)\big) =\prod_{i=1}^{K} p(x_i)$}, where $p(x_i)$ is the capacity achieving distribution of the sub-channels, shown in \cite[Lemma~3]{lenz_upper_2019} to be the uniform distribution. Each of the multi-draw channels $i$, $1 \leq i \leq K$, has the capacity $C _{d_i}$. Since the transmissions over the $K$ parallel channels are independent of each other, we can conclude for any given, fixed vector $\d = (d_1,\dots,d_K)$
\begin{align*}
\mathrm{C}_\d &= \frac{1}{K} \underset{p(x_1,\ldots,x_K)}{\max} \ I (X_1,\ldots,X_K;\cY_1,\ldots,\cY_K) \\
&= \frac{1}{K}\sum_{i=1} ^K \underset{p(x_i)}{\max} \ I (X_i;\cY_i) = \frac{1}{K}\sum_{i=1} ^K  \mathrm{C}_{d_i}.
\end{align*}

This means, that for a fixed and known $\d$, $C_\d$ is clearly achievable. However, here we aim to design a codebook that achieves vanishing error probability for all $\d \in \dplussetmax := \{\d \in \NN_0^K: \sum _{i=1} ^K d_i \leq \dpmax\}$ and $R<C_\d$. Note that $\dplussetmax$ has finite cardinality resulting in a total number of $\lvert \dplussetmax \rvert$ possible channels for vectors $\d$. 

As a consequence we can use similar arguments as in \cite{willems_signaling_2008} with a Cover-like random coding argument to show that the capacity expression is achievable. We design the codebook $\mathcal{X} := \{ \mathbf{X}(w), w \in [2^{RK\ell}]\}$, where each codeword $\mathbf{X}(w)$ with message index $w$ is a matrix of size $K\times \ell$. Note that the design of this codebook only depends on $\dplussetmax$ and not the specific channel realization as desired above. Now, given $\d \in \dplussetmax$, for each $K$-parallel multi-draw channel with draws $\d$, there exists a decoder $\widehat{w} _\d$, $\d \in \dplussetmax$ that has error probability $\Pr(\widehat{w}_\d\neq w)<\epsilon$, if $R<C_\d-\epsilon$ by the standard joint typicality argument using the mutual information rate derived above. Therefore, choosing a code $\mathcal{X}$ of rate $R < \mathrm{C}_\d-\epsilon$, for $\ell$ sufficiently large, by a union bound argument
\begin{align*}
\bar{\Pr}\left(\bigcup_{\substack{\d \in \dplussetmax:\\R<C_\d-\epsilon}}\widehat{w}_\d\neq w\right) \leq \lvert \dplussetmax \rvert \epsilon.
\end{align*}
Since $|\dplussetmax|$ is finite we can choose $\epsilon$ as small as desired, any rate $R < C_\d$ is achievable and thus there exist a codebook with vanishing error probability over all channels. 
\end{proof}

\begin{remark}
Note that decoding a random code, as in the proof of Lemma~\ref{lem:inner_capacity_achieving_code}, still involves a complexity that is only polynomial in $M$. However, more practical capacity-achieving code families for the $K$-parallel multi-draw channel might be obtained by \cite{willems_signaling_2008}'s rate-matching codes or \cite{hof_capacity-achieving_2013}'s polar code construction.
\end{remark}

In Lemma~\ref{lem:inner_capacity_achieving_code} above, we have seen that there are inner codes that achieve the capacity of several $K$-parallel multi-draw channels at the same time. A key towards deriving an achievable outer rate for a given inner code is thus to count the number of blocks in which the draw vectors $\d$ fall within the set of channels whose capacity is achieved (i.e., how many inner blocks can be recovered with high probability). The following lemma is a generalization of \cite[Lemma~2]{lenz_upper_2019} from $K=1$ to blocks of size $K$ and states that, asymptotically, the relative number of blocks with draw vector $\d$ is with high probability close to the pmf of a $K$-variate Poisson distribution with parameter $c$.

%

%

\begin{lemma} \label{lem:relative_number_of_draws_Poisson}

	Let $\epsilon>0$ and $c \geq 1$ be fixed. Then, for $N=cM$ and $M$ sufficiently large, 
	\begin{equation*}
	\Pr\Big( \sum _{\d \in \NN_0^K} \left| Q_\d - \tfrac{M}{K}p_c(\d) \right| < \epsilon M \Big) > 1-\epsilon.
	\end{equation*}
\end{lemma}


\begin{proof}
In the beginning we need to derive expressions for $\Exp [Q_\d]$ and $\Var[Q_\d]$, which we need later to formally prove the statement of the lemma via Chebyshev's inequality.
	
Let $\dset$ be the set of all possible combinations for a vector $\d$ defined as
$\dset := \{\d \in \NN_0^K \, : \, \sum_{i=1}^{K} d_i \leq cM\}$ with $M \in \ZZ$ such that $K \mid M$ and furthermore, $ \dplusset:= \{ \d \in \NN_0^K: \, \sum_{i=1}^{K} d_i = d\} \subseteq \dset$ for some fixed $d \in \{0,1,\ldots, N\}$.

We define the indicator function $I_\d(i)$ and an event for all $i = 1, \ldots, \nicefrac{M}{K}$ blocks in the following:
\begin{align*}
	I_\d(i) &:= \begin{cases}
	1, &\text{if } D_{K(i-1)+j}=d_j \text{ for all } j=1,\dots,K, \\
	0, &\text{else};
	\end{cases} \\
	\dpevent &:= \{ (D_1, \dots,D_M): \, \sum ^K _{j=1} D_{K(i-1)+j} = d \}.
\end{align*}
We can thus compute the probability
\begin{align*}
	&\Pr \left( I_\d(i) = 1\right) 
	= \Pr\big( (D_1,\dots,D_K) = (d_1,\dots,d_K) \big) \\
	&= \Pr\big( (D_1,\dots,D_K) = (d_1,\dots,d_K) \big) \mid \dpeventplus\big)  \cdot \Pr \left( \dpeventplus \right),
\end{align*}
where we abbreviate $\dplus = \sum_{i=1}^Kd_i$. In the last equation the second term can be interpreted as the probability of drawing exactly $\dplus$ strands in one block of length $K$. Thus, we obtain
\begin{align*}
	\Pr \left( \dpeventplus \right) = \tbinom{N}{d}\left(\frac{1}{\nicefrac{M}{K}}\right) ^\dplus \left( 1- \frac{1}{\nicefrac{M}{K}} \right) ^{N-\dplus},
\end{align*}
since there are in total $\nicefrac{M}{K}$ blocks. 
For clarity, we abbreviate the first term by
\begin{align*}
	&\multidp :=  \Pr \left( D_j = d_j \mid \dpeventplus,  \, \forall \, j=1,\dots,K \right).
\end{align*}
	
Observe that we can express $Q_\d$ as the sum of the indicator variables, i.e., $Q_\d = \sum_{i=1}^{\nicefrac{M}{K}} I_\d(i)$. We compute 
\begin{align*}
	&\Exp[Q_\d] =  \sum_{i=1}^{\nicefrac{M}{K}} \Exp[I_\d(i)]= \sum_{i=1}^{\nicefrac{M}{K}} \Pr\left( I_\d(i) = 1\right) \\
	&= \frac{M}{K}\cdot\tbinom{N}{\dplus}\left(\frac{1}{\nicefrac{M}{K}}\right) ^\dplus \left( 1- \frac{1}{\nicefrac{M}{K}} \right) ^{N-\dplus} \cdot \multidp \\
	&\leq \dfrac{M}{K} \dfrac{N^\dplus}{\dplus ! } \dfrac{1}{\left( \nicefrac{M}{K} \right) ^\dplus } e^{-\frac{KN}{M}} e^{\frac{\dplus K}{M}} \multidp \\
	&= \dfrac{M}{K} p_{Kc} (\dplus)  e^{\frac{\dplus K}{M}} \multidp,
\end{align*}
where we used for the inequalities that $1-x \leq e^{-x}$ for any $x \in \RR$ and $\binom{N}{\dplus} \leq \frac{N^\dplus}{\dplus !}$ for any $N, \dplus \in \NN_0$ with $N \geq \dplus$. Moreover, 
{\small
\begin{align*}
	&\Exp[Q_\d ^2] 
	= \Exp[Q_\d] + \multidp ^2 \\
	&\qquad \cdot \dfrac{M}{K} \left( \dfrac{M}{K} -1 \right) \dfrac{N^{[2\dplus]}}{(\dplus !)^2} \left(\frac{1}{\nicefrac{M}{K}}\right) ^{2\dplus} \left( 1- \frac{2}{\nicefrac{M}{K}} \right) ^{N-2\dplus}. 
\end{align*}}
In the last equality of the equation before we used $N^{[2\dplus]} = N (N-1) \ldots (N-2 \dplus + 1)$. Using the aforementioned bounds and additionally using that $(1-x)^{\dplus} \geq 1 - \dplus x$ for any $0 < x < 1$ and $\dplus \in \NN_0$ and $\dplus \cdot p_{Kc}(\dplus) \leq Kc$, we obtain the following upper bound after some manipulations.
\begin{align*}
	\Var[Q_\d] &\leq \dfrac{M}{K} p_{Kc} (\dplus) e^{\frac{\dplus K}{M}} \multidp \\
	&\quad + \left( \frac{M}{K} \right) ^2 p_{Kc} ^2(\dplus)  \left(1-\left(1-\frac{2}{\nicefrac{M}{K}} \right) ^\dplus \right) \\
	&\qquad \cdot e^{\frac{4\dplus K}{M}} \multidp ^2 \\
	&\leq \dfrac{M}{K} p_{Kc} (\dplus) e^{4 Kc} \multidp (1+2Kc)
	\end{align*}
	
Note that this derivation extends the results from \cite{kolchin1978random} to arbitrary block sizes $K$.
	
With these valid expressions for $\Exp [Q_\d]$ and $\Var[Q_\d]$ we are prepared to prove the lemma after minor definitions.
We abbreviate the statement of the lemma by
$$\mathcal{Q} = \left\{ \Q: \sum _{\d \in \dset} \left| Q_\d - \tfrac{M}{K}p_c(\d) \right| < \epsilon M \right\}.$$
%
%
Let $b_{M}(\d)$ be a non-negative real number such that the sum $\sum_{\d \in \dset} b_{M}(\d) \leq \epsilon M$ for large enough $M$ in the following. Hence, we can derive via the triangular inequality for $|Q_\d -\frac{M}{K}p_c(\d)| \leq |Q_\d - \Exp[Q_\d]|+| \Exp[Q_\d] -\frac{M}{K}p_c(\d)|$ and Chebyshev's inequality
\begin{align*}
&\Pr (\mathcal{Q}) = \Pr\Big( \sum _{\d \in \dset} \left| Q_\d - \tfrac{M}{K}p_c(\d) \right| < \epsilon M \Big) \\
&= 1-\Pr\Big( \sum _{\d \in \dset} \left| Q_\d - \tfrac{M}{K}p_c(\d) \right| \geq \epsilon M \Big) \\
&\overset{(a)}{\geq} 1-\sum _{\d \in \dset}\Pr\Big( \left| Q_\d - \tfrac{M}{K}p_c(\d) \right| \geq b_M(\d)\Big)\\
&\geq 1- \sum _{\d \in \dset}\Pr \Big( |Q_\d - \Exp[Q_\d]| \geq b_M(\d) -| \Exp[Q_\d] -\tfrac{M}{K}p_c(\d)| \Big) \\
&> 1 - \sum _{\d \in \dset}\frac{\Var [Q_\d]}{ \left( b_M(\d) - \left| \Exp[Q_\d] - \frac{M}{K}p_c(\d) \right|\right)^2},
\end{align*}
where we used a union bound argument for all $\d \in \dset$ in $(a)$. 
	%
We now choose $b_M(\d)$ as
$$b_M(\d) = | \Exp[Q_\d] - \tfrac{M}{K} p_c(\d) | + \sqrt{M} M^{\epspr} \sqrt{p_{Kc} (\dplus) } \; \dplus.$$
Observe that one can verify that $p_c(\d) = p_{Kc}(\dplus) \cdot \multidp$. Inserting this choice of $b_M(\d)$ into the expression of $\Pr(\mathcal{Q})$ we get 
%
\begin{align*}
&\Pr (\mathcal{Q}) > 1 - M^{-2\epspr} \frac{e^{4Kc} (1+2Kc) }{K} \sum_{\d \in \dset} \frac{\multidp}{\dplus^2}\\
&=1 - M^{-2\epspr} \frac{e^{4Kc} (1+2Kc) }{K} \sum_{d=0}^N \frac{1}{d^2} \sum_{\d \in \dplusset} B(d,\d)
\end{align*}
Due to $\sum _{\d \in \dplusset} B(d,\d) = 1$ it follows that for any $\epspr > 0$ and large enough $M$ this will result to $\Pr (\mathcal{Q}) > 1 - \epsilon$.

It remains to prove the initial assumption that
{
	\begin{align*}
	\sum _{\d \in \dset}b_M(\d)	< \epsilon M
	\end{align*}
}
for large enough $M$. For the sake of brevity we omit to show that $\sum _{\d \in \dset} \left| \Exp[Q_\d] - \frac{M}{K} p_c(\d)\right| < \tfrac{\epsilon}{2} M$ and refer as a guidance of the proof to \cite{lenz_upper_2019}. 


Let us denote $\Delta _{M}(\d)= \sqrt{M} M^{\epspr} \sqrt{p_{Kc} (\dplus) } \; \dplus$ as the remaining summand in $b_M(\d)$ and investigate its sum as follows.
\begin{align*}
&\sum _{\d \in \dset} \Delta_{M}(\d) = M^{\frac{1}{2}+\epspr} \sum _{d = 0} ^N  \sqrt{p_{Kc} (d) } \; d \sum _{\d \in \dplusset} 1  \\
&\overset{(b)}{\leq}  M^{\frac{1}{2}+\epspr} \sum _{d = 0} ^N  e^{\frac{Kc}{2}} \sqrt{Kc} p_{Kc} \left(\left\lfloor \frac{d}{2} \right\rfloor \right) d ^{(K+1)} \\
&\overset{(c)}{\leq}  M^{\frac{1}{2}+\epspr} e^{\frac{Kc}{2}} \sqrt{Kc} (2 \cdot 3^{K+1}) \sum _{d = 0} ^{\lceil \frac{N}{2} \rceil }   p_{Kc} (d) \dplus  ^{(K+1)} \\
&\overset{(d)}{\leq}  M^{\frac{1}{2}+\epspr} e^{\frac{Kc}{2}} \sqrt{Kc} (2 \cdot 3^{K+1}) \sum _{d = 0} ^{ \infty }   p_{Kc} (d) d  ^{(K+1)} 
< \tfrac{\epsilon}{2} M
\end{align*}
Here we used for $(b)$ that it holds $|\dplusset|  \leq d ^K$. Additionally, the inequality $\sqrt{p_{Kc} (d) } \leq e^{\frac{Kc}{2}} \sqrt{Kc} p_{Kc} (\lfloor \frac{d}{2} \rfloor)$ is used, where $\sqrt{d!} > \lfloor \frac{d}{2} \rfloor ! $ is applied and due to the flooring operation it holds $(Kc)^{\frac{d}{2}} \leq \sqrt{Kc} (Kc)^{\lfloor \frac{d}{2} \rfloor}$ both for $d \in \NN_0$ and $K,c \geq 1$. For $(c)$ we have split the sum in odd and even terms but directly used the upper bound $(2 d) ^{(K+1)} + (2d + 1) ^{(K+1)} \leq 2 \cdot (3d)^{K+1}$ to combine the split terms again. In the step $(d)$ we transform the expression such that it is resulting to the $K+1$-th moment expression $\sum _{d = 0} ^{ \infty } p_{Kc} (d) d  ^{(K+1)}$ of the Poisson distribution with parameter $Kc$ and $d$. Any moment of the Poisson distribution is finite and therefore we can conclude the last inequality for $M$ sufficiently large and small $\epspr > 0$. 

As a consequence both terms of $b_M(\d)$ are not greater than $\tfrac{\epsilon}{2} M$, which concludes the proof.
\end{proof}

The previous two lemmas have answered the questions under which conditions an inner block can be recovered given that all of its strands are properly clustered and indices are correctly retrieved. The next and last lemma shows that asymptotically, most strands are indeed correctly clustered and index-recovered with high probability under the assumption of Lemma \ref{lem:relative_number_of_draws_Poisson}. 


\begin{lemma} \label{lemma:wrong:clusters:decoding}
	Let $\PoiE$ be the event from Lemma \ref{lem:relative_number_of_draws_Poisson}. Fix \mbox{$\epsilon>0$}, $\beta<{\Rin(1-\nicefrac{\beta}{\Rix})(1-h(2p))}$. Then
	%
	for all \mbox{$\Q \in \PoiE$}, 
	\begin{align*}
		\Pr(\MC > \epsilon M |\Q ) &\leq \epsilon, \\
		\Pr(\MIx > \epsilon M | \Q ) &\leq \epsilon,\\
		\Pr(\MIn > \epsilon M | \Q ) &\leq \epsilon
	\end{align*}
	for large enough $M$.
\end{lemma}

\begin{proof}
	Let $\epsilon'>0$ be arbitrary and $J_j$ be the indicator that is $1$, if $d(\y_j,\x_{I_j}) >(p+\tfrac{\epsilon'}{2})L$ or if there exists an output strand $j' \in [N]$, $I_j\neq I_{j'}$ with distance $d(\y_j,\y_{j'}) \leq \rho L$, where $\rho L$ with $\rho = 2p+\epsilon'$ is the diameter of the clustering algorithm introduced in Section \ref{ssec:decoding}. Then the number of wrong clusters is at most $\MC \leq \sum_{j=1}^{N} J_j$. The individual probabilities can be bounded by
	\begin{align*}
	\Pr(d(\y_j,\x_{I_j} )>(p+\tfrac{\epsilon'}{2})L) &= \sum_{i=(p+\tfrac{\epsilon'}{2})L}^{L} \binom{L}{i}p^i(1-p)^{L-i} \\
	& \overset{(a)}{\leq} \mathrm{e}^{-LD(p+\tfrac{\epsilon'}{2}||p)} \leq \epsilon'/2
	\end{align*}
	for large enough $L$, where $D(a||p)$ is the Kullback-Leibler divergence between Bernoulli variables with success probabilities $a$ and $p$. Here we used the Chernoff bound on the binomial tail in inequality $(a)$. On the other hand, the first $\Rin \ell$ bits of each strand $\y_j$ form a uniform random vector, as the marginal distribution of any symbol in a MDS code is uniform. It follows that we can bound the probability of another strand being within the clustering diameter
	\begin{align*}
		\Pr&(\exists j':I_j\neq I_{j'}, d(\y_j,\y_{j'})\leq \rho L) \\
		& \overset{(b)}{\leq} N \Pr(d(\y_j,\y_{j'})\leq \rho \Rin \ell) \leq N2^{-\Rin\ell}\sum_{i=0}^{\rho\Rin\ell} \binom{\Rin\ell}{i} \\
		& \overset{(c)}{\leq} c 2^{\beta L} 2^{\Rin\ell (h(2p+\epsilon')-1)} \leq \epsilon'/2
	\end{align*}
	for large enough $L$. Here, $(b)$ follows from the union bound over all output strands and $(c)$ follows from the Chernoff bound on binomial tails, where $h(\bullet)$ denotes the binary entropy function. Thus $\Pr(J_j = 1) \leq 2\epsilon'$ for large enough $L$. It follows that $ \Exp(\sum_{j=1}^NJ_j) \leq \epsilon' N$ and $ \Var(\sum_{j=1}^NJ_j) \leq \epsilon' N + \epsilon' N^2 $. The statement then follows from Chebyshev's inequality.
	
	The second claim follows from the fact that for any $\epsilon'>0$ the probability of erroneous index decoding is at most $\epsilon'$ for any cluster with $C_{d_i} \geq \Rix$ and large enough $L$ as we are decoding a capacity achieving code with rate less than the capacity. Given $\Q$, the individual decoding processes are statistically independent and thus $\MIx$ can be upper bounded by a binomial variable with success probability $\epsilon'$ and $N$ trials. Hence,
	$$ \Pr(\MIx>\epsilon M) \leq \sum_{i=\epsilon M}^{N} \binom{N}{i} {(\epsilon')}^i(1-\epsilon')^{N-i}  \overset{(a)}{\leq} \mathrm{e}^{-ND(c\epsilon||\epsilon')},  $$
	where $(a)$ follows from the Chernoff bound. Hence $\Pr(\MIx>\epsilon M)$ is at most $\epsilon$ for large enough $M$, respectively $N$.
	
	The bound for $\MIn$ is obtained by choosing a $\dpmax$ such that
	$$ \sum_{\dplus > \dpmax} p_{Kc}(\dplus) < \epsilon'. $$
	Then, for all blocks for which $d_{bK+1}+\dots+d_{bK+K} \leq \dpmax$ the number of errors can be bounded as for the case of $\MIx$ using the bound on the error probability of inner decoding in Lemma \ref{lem:inner_capacity_achieving_code}. Since the number of blocks with $d_{bK+1}+\dots+d_{bK+K} > \dpmax$ is given by $Q_{\dpmax+1}+Q_{\dpmax+1}+\dots$, which is at most $M (\sum_{\dplus>\dpmax}p_{Kc}(\dplus)+\epsilon') \leq 2M\epsilon'$ with high probability (c.f. Lemma \ref{lem:relative_number_of_draws_Poisson}), the claim follows.
\end{proof}

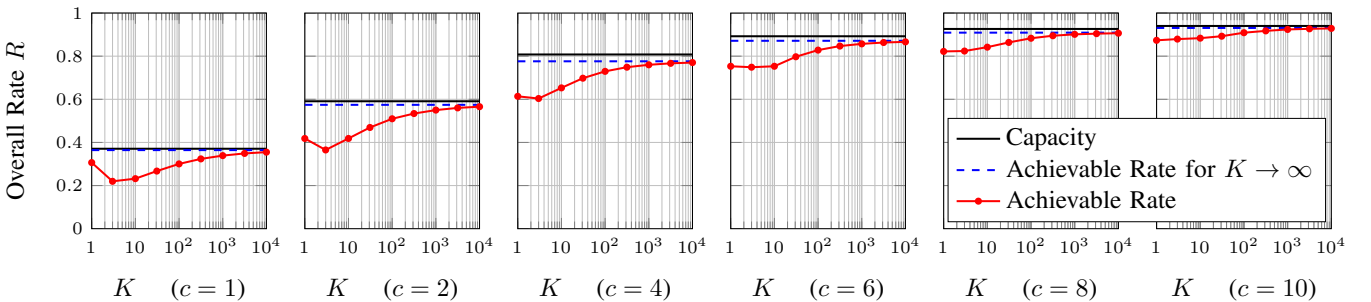
\begin{figure*}[b]
\resizebox{\textwidth}{!}{
\begin{tikzpicture}
\pgfplotsset{
  compat = 1.3,
  height=3cm,
  scale only axis,
  xmajorgrids,
  xminorgrids,
  ymajorgrids,
  yminorgrids,
  every axis x label/.style={at={(0.5,0)}, align=center, yshift=-24pt},
  every axis y label/.style={at={(0,0.5)}, align=center, xshift=-30pt, rotate=90},
  legend pos = south east,
  legend cell align=left,
}

\begin{groupplot}[
  group style={group size=6 by 1,
  horizontal sep=15pt,
  vertical sep=15pt}
]

	\nextgroupplot[
	  width=0.133333\textwidth,
	  xlabel={$K$ \quad ($c=1$)},
	  ylabel={Overall Rate $R$},
	  xmin=1,
	  xmax=10000,
	  xtick={1,10,100,1000,10000,100000},
      xticklabels={$1^{\vphantom{5}}$,$10^{\vphantom{5}}$,$10^2$,$10^3$,$10^4$,$10^5$},
	  xticklabel style = {font=\scriptsize},
	  yticklabel style = {font=\scriptsize},
	  ymin=0,
	  ymax=1,
	  title={},
	  xmode=log,
	]

\addplot [solid, color=black, thick] table[row sep=\\] {
1 0.370762 \\
10000 0.370762 \\
};

\addplot [dashed, color=blue, thick] table[row sep=\\] {
1 0.364443 \\
10000 0.364443 \\
};

\addplot [solid, color=red, thick, mark=*, mark size=1pt] table[row sep=\\] {
1.000000 0.306575 \\
3.000000 0.219713 \\
10.000000 0.232054 \\
31.000000 0.267227 \\
100.000000 0.300439 \\
316.000000 0.323685 \\
1000.000000 0.338902 \\
3162.000000 0.349049 \\
10000.000000 0.354996 \\
};

	\nextgroupplot[
	  width=0.133333\textwidth,
	  xlabel={$K$ \quad($c=2$)},
	  ylabel={},
	  yticklabels={},
	  xmin=1,
	  xmax=10000,
	  xtick={1,10,100,1000,10000,100000},
      xticklabels={$1^{\vphantom{5}}$,$10^{\vphantom{5}}$,$10^2$,$10^3$,$10^4$,$10^5$},
	  xticklabel style = {font=\scriptsize},
	  ymin=0,
	  ymax=1,
	  title={},
	  xmode=log,
	]

\addplot [solid, color=black, thick] table[row sep=\\] {
1 0.590899 \\
10000 0.590899 \\
};

\addplot [dashed, color=blue, thick] table[row sep=\\] {
1 0.574362 \\
10000 0.574362 \\
};

\addplot [solid, color=red, thick, mark=*, mark size=1pt] table[row sep=\\] {
1.000000 0.418345 \\
3.000000 0.365103 \\
10.000000 0.418443 \\
31.000000 0.469370 \\
100.000000 0.509697 \\
316.000000 0.533934 \\
1000.000000 0.549960 \\
3162.000000 0.560151 \\
10000.000000 0.565750 \\
};

	\nextgroupplot[
	  width=0.133333\textwidth,
	  xlabel={$K$ \quad($c=4$)},
	  ylabel={},
	  yticklabels={},
	  xmin=1,
	  xmax=10000,
	  xtick={1,10,100,1000,10000,100000},
      xticklabels={$1^{\vphantom{5}}$,$10^{\vphantom{5}}$,$10^2$,$10^3$,$10^4$,$10^5$},
	  xticklabel style = {font=\scriptsize},
	  ymin=0,
	  ymax=1,
	  title={},
	  xmode=log,
	]

\addplot [solid, color=black, thick] table[row sep=\\] {
1 0.807848 \\
10000 0.807848 \\
};

\addplot [dashed, color=blue, thick] table[row sep=\\] {
1 0.776161 \\
10000 0.776161 \\
};

\addplot [solid, color=red, thick, mark=*, mark size=1pt] table[row sep=\\] {
1.000000 0.613783 \\
3.000000 0.603571 \\
10.000000 0.652942 \\
31.000000 0.697839 \\
100.000000 0.729575 \\
316.000000 0.748908 \\
1000.000000 0.759920 \\
3162.000000 0.766720 \\
10000.000000 0.770345 \\
};

	\nextgroupplot[
	  width=0.133333\textwidth,
	  xlabel={$K$ \quad($c=6$)},
	  ylabel={},
	  yticklabels={},
	  xmin=1,
	  xmax=10000,
	  xtick={1,10,100,1000,10000,100000},
      xticklabels={$1^{\vphantom{5}}$,$10^{\vphantom{5}}$,$10^2$,$10^3$,$10^4$,$10^5$},
	  xticklabel style = {font=\scriptsize},
	  ymin=0,
	  ymax=1,
	  title={},
	  xmode=log,
	]

\addplot [solid, color=black, thick] table[row sep=\\] {
1 0.892294 \\
10000 0.892294 \\
};

\addplot [dashed, color=blue, thick] table[row sep=\\] {
1 0.871261 \\
10000 0.871261 \\
};

\addplot [solid, color=red, thick, mark=*, mark size=1pt] table[row sep=\\] {
1.000000 0.753098 \\
3.000000 0.748783 \\
10.000000 0.753155 \\
31.000000 0.797214 \\
100.000000 0.828103 \\
316.000000 0.846844 \\
1000.000000 0.857006 \\
3162.000000 0.863287 \\
10000.000000 0.866277 \\
};

	\nextgroupplot[
	  width=0.133333\textwidth,
	  xlabel={$K$ \quad($c=8$)},
	  ylabel={},
	  yticklabels={},
	  xmin=1,
	  xmax=10000,
	  xtick={1,10,100,1000,10000,100000},
      xticklabels={$1^{\vphantom{5}}$,$10^{\vphantom{5}}$,$10^2$,$10^3$,$10^4$,$10^5$},
	  xticklabel style = {font=\scriptsize},
	  ymin=0,
	  ymax=1,
	  title={},
	  xmode=log,
	]

\addplot [solid, color=black, thick] table[row sep=\\] {
1 0.926173 \\
10000 0.926173 \\
};

\addplot [dashed, color=blue, thick] table[row sep=\\] {
1 0.908990 \\
10000 0.908990 \\
};

\addplot [solid, color=red, thick, mark=*, mark size=1pt] table[row sep=\\] {
1.000000 0.822000 \\
3.000000 0.823735 \\
10.000000 0.841766 \\
31.000000 0.863660 \\
100.000000 0.883046 \\
316.000000 0.894715 \\
1000.000000 0.901209 \\
3162.000000 0.904369 \\
10000.000000 0.906417 \\
};

	\nextgroupplot[
	  width=0.133333\textwidth,
	  xlabel={$K$ \quad($c=10$)},
	  ylabel={},
	  yticklabels={},
	  xmin=1,
	  xmax=10000,
	  xtick={1,10,100,1000,10000,100000},
      xticklabels={$1^{\vphantom{5}}$,$10^{\vphantom{5}}$,$10^2$,$10^3$,$10^4$,$10^5$},
	  xticklabel style = {font=\scriptsize},
	  ymin=0,
	  ymax=1,
	  title={},
	  xmode=log,
	]

\addplot [solid, color=black, thick] table[row sep=\\] {
1 0.940040 \\
10000 0.940040 \\
};

\addlegendentry{Capacity};

\addplot [dashed, color=blue, thick] table[row sep=\\] {
1 0.930767 \\
10000 0.930767 \\
};

\addlegendentry{Achievable Rate for {$K \to \infty$}};

\addplot [solid, color=red, thick, mark=*, mark size=1pt] table[row sep=\\] {
1.000000 0.873463 \\
3.000000 0.879157 \\
10.000000 0.883433 \\
31.000000 0.892538 \\
100.000000 0.908783 \\
316.000000 0.917221 \\
1000.000000 0.923865 \\
3162.000000 0.926786 \\
10000.000000 0.928672 \\
};

\addlegendentry{Achievable Rate};

\end{groupplot}
\end{tikzpicture}
}
\vspace{-0.7cm}
\caption{Overall rate $R$ over block size $K$ (logarithmic scale) for varying channel parameter $c=1,2,4,6,8,10$. Fixed parameters: $\beta = \nicefrac{1}{20}$ and $p=0.1$. For each $c$, the index rate is chosen to maximize the achievable rate for $K \to \infty$ given in Corollary~\ref{cor:achievable_rate_large_k}: $\Rix = 0.999 \cdot C_{d^*}$ with $d^*=1$ for $c=1,2,4$, $d^*=2$ for $c=6,8$, and $d^*=3$ for $c=10$. The curves for finite $K$ are obtained by Monte-Carlo simulations with $N=10^4$ samples.}
\label{fig:R_over_k_joined}
\end{figure*}


\subsection{Proof of Main Statement}\label{ssec:proof_of_main_statement}


\begin{proof}[Proof of Theorem~\ref{thm:achievability}]
We will prove the theorem by showing that the probability of a decoding error inside the outer codeword goes to zero as $M \to \infty$. In this view, let $s,t$ denote the random variables that count number of erasures, respectively errors inside the outer codeword after clustering, index decoding and inner decoding. Recalling the definitions of the number of wrong clusters $\MC$, and wrongly decoded indices $\MIx$, respectively inner codewords $\MIn$ from Section \ref{ssec:decoding}, we obtain 
\begin{align*}
s &\leq \textstyle\sum_{\d:C_\d(\Rix)\leq \Rin} Q_\d +2M_\mathrm{C}+ 2\MIx,  \\
t &\leq \MIn+2M_\mathrm{C}+ 2\MIx .
\end{align*}
This can be explained as follows. First, the inner decoder declares an erasure, when $C_{\widehat{\d}} \leq \Rin$ for a block with $\widehat{\d}$ draws. Assuming no clustering errors or index decoding errors, the first term counts the number of erasures declared by the inner decoder. Then, for each wrong cluster or a wrongly decoded index, there can be at most two inner codewords affected - the inner codeword, where the cluster actually belongs to and the inner codeword that the cluster is mistakenly associated to. The same argument can be used for the upper bound on $t$. We obtain for the success probability
\begin{align*}
\Pr(\SucE) &= \textstyle\sum_\Q \Pr(\SucE|\Q) \Pr(\Q)  \\
&\overset{(a)}{\geq} \textstyle\sum_{\Q \in \PoiE}\Pr(s+2t\leq M(1-\Rout)|\Q)\Pr(\Q),
\end{align*}
where $(a)$ holds by the properties of unique decoding of MDS codes. We further obtain for the conditional success probability
\begin{align*}
&\Pr(s+2t\leq M(1-\Rout)|\Q)  \\
&=1- \Pr(s+2t+10\epsilon M> M(1-\Rout)+ 10\epsilon M|\Q)   \\
 &\geq 1- \Pr(s > M(1-\Rout)-10\epsilon M|\Q) - \Pr(t > 5\epsilon M |\Q).
\end{align*}

On the one hand, we have
{\small\begin{align*}
	&\Pr(t > 5\epsilon M |\Q) \leq \Pr(\MIn+2M_\mathrm{C}+ 2\MIx > 5 \epsilon M|\Q) \\
	 &\leq \Pr(\MC\!>\! \epsilon M|\Q) \!+\! \Pr(\MIx\!>\! \epsilon M|\Q)
	\!+\! \Pr(\MIn\!>\! \epsilon M|\Q) \overset{(b)}{\leq} 3\epsilon,
\end{align*}}%
where $(b)$ follows from Lemma \ref{lemma:wrong:clusters:decoding}. On the other hand, abbreviate $\mathcal{D}_C= \{\d \in \NN_0^K:C_\d(\Rix) \leq \Rin\}$. Then, for $\Q \in \PoiE$
{\small
\begin{align*}
	&\Pr(s > M(1-\Rout)-10\epsilon M|\Q) \\
	&\leq \Pr \Big(\sum_{\d\in \mathcal{D}_C} Q_\d+2M_\mathrm{C}+ 2\MIx> M(1-\Rout)-10\epsilon M|\Q\Big) \\
	& \overset{(c)}{\leq} \Pr\Big(2M_\mathrm{C}+ 2\MIx> M(1-\Rout-\sum_{\d\in \mathcal{D}_C} p_c(\d)-11\epsilon )|\Q\Big) \\
	& \overset{(d)}{\leq} \Pr\left(2M_\mathrm{C}+ 2\MIx> 4 \epsilon M|\Q\right) \leq 2 \epsilon
\end{align*}
}
where in $(c)$ we used that $\Q \in \PoiE$ and in $(d)$, we chose $\Rout = \sum _{\d: {C}_\d(\Rix) > \Rin} p_c(\d) - 15\epsilon$. Finally inserting these results into the overall success probability, and using the bound on $\Pr(\PoiE)$ from Lemma \ref{lem:relative_number_of_draws_Poisson}, we obtain
\begin{align*}
\Pr(\SucE) \geq (1-5\epsilon) \sum_{\Q \in \PoiE}\Pr(\Q)  \geq (1-5\epsilon)(1-\epsilon)
\end{align*}
for large enough $M$ and the claim of the theorem follows, as we can choose $\epsilon$ as small as desired.
\end{proof}	

\section{Numerical Results}\label{sec:numerical_results}

\shortlongversion{The achievable rate given in the main statement of the paper (Theorem~\ref{thm:achievability}) is a semi-closed expression and can be efficiently approximated by Monte-Carlo simulation.
}{The achievable rate given in the main statement of the paper (Theorem~\ref{thm:achievability}) is a semi-closed expression it is not obvious from the formulation how to compute it efficiently.

One way is to approximate it by summing only over a finite subset $\mathcal{A} \subset \NN_0^K$ of vectors $\d$ for which $\sum_{\d \in \NN_0^K \setminus \mathcal{A}} p_c(\d)$ is sufficiently small (e.g., $\mathcal{A} = \{\d \in \NN_0^K : \sum_{i=1}^{K} d_i \leq \dplus\}$ for a large $\dplus$).
A drawback of this approach is that the cardinality of the set $\mathcal{A}$ might grow exponentially\footnote{For instance, in the case $\mathcal{A} = \{\d \in \NN_0^K : \sum_{i=1}^{K} d_i \leq \dplus\}$, the cardinality is related to the number of partitions of integers $\leq \dplus$.} in the channel parameters $K,c$ and the inverse of the sought precision. A second and much more efficient approach is to approximate the achievable rate by a Monte-Carlo simulation.
}
The foundation of this method is
that the achievable rate equals the probability $\Pr\big(\tfrac{1}{K} \sum_{i=1}^K C_{d_i}(\Rix)>\Rin \big)$ for a random variable $\d$ drawn from a $K$-variate Poisson distribution with parameter $c$.
Hence, we can repeatedly take samples of $\d$ and estimate the achievable rate by the relative number of times that $\tfrac{1}{K} \sum_{i=1}^K C_{d_i}(\Rix)>\Rin$.
The complexity of this method grows only linearly in $K$ and the inverse of the sought variance of the estimator.

\shortlongversion{}{In the following, we give numerical results obtained from the Monte-Carlo method.}
All simulations were performed using the computer-algebra system SageMath \cite{sagemath} (version~8.1). As fixed channel parameters, we took $\beta=\nicefrac{1}{20}$ and $p=0.1$, which are common values for nanopore sequencing, see e.g.~\cite{heckel_characterization_2019,lenz_coding_2019}.

Figure~\ref{fig:R_over_k_joined} presents the achievable overall rate $R$ as a function of the block length $K$ for several channel parameters $c$. For comparison, it displays the capacity of the channel \cite{lenz_upper_2019,lenz_achieving_2020} and the achievable rate for $K \to \infty$ (cf.~Corollary~\ref{cor:achievable_rate_large_k}).
In all cases, the achievable rate for $K \to \infty$ is close to the capacity. 
Furthermore, for all simulated parameters, the achievable rate approaches the expected asymptotic value for growing $K$.
There is a significant gap between the asymptotic rate and the rate for $K=1$, especially for $c=2,4,6$.
For small $c$, the achievable rate first decreases for small $K$, but then increases until close to the asymptotic value.
Using a sufficiently large block length $K$ (e.g., $K=100$ for $c\geq 2$), the achievable rate is significantly larger than for $K=1$, and can get very close to the capacity.

\begin{figure}[tb]
\input{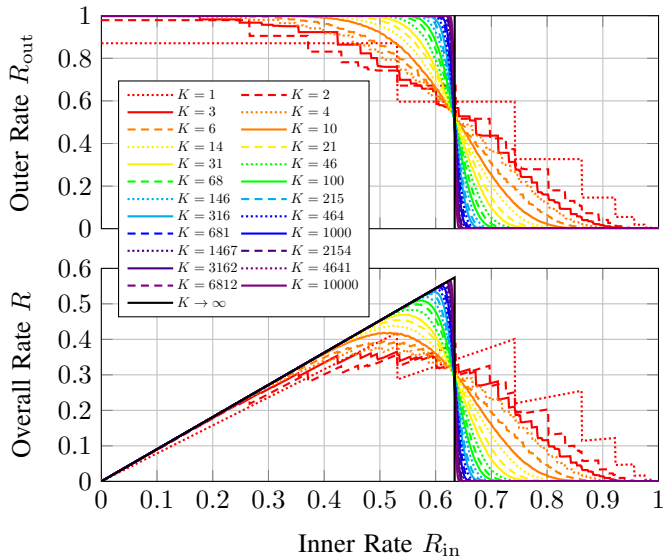}
\vspace{-0.7cm}
\caption{Achievable outer rate $R_\mathrm{out}$ and overall rate $R$, respectively, plotted over the inner rate $R_\mathrm{in}$ for several $K$ (here: $K=\lfloor 10^{\nicefrac{i}{6}} \rfloor$ for $i=1,\dots,24$). Fixed parameters: $c=2$, $\beta=\nicefrac{1}{20}$, $p=0.1$, $R_\mathrm{ix} = 0.5304$ ($=0.999 \cdot C_1$, where $C_1$ is the capacity of the multi-draw channel with $d=1$ draw and crossover probability $p$). The curves for finite $K$ are obtained from Monte-Carlo simulations with $N=10^4$ samples. The curves for $K \to \infty$ show the theoretical predictions obtained from Lemma~\ref{cor:achievable_rate_large_k}.
}
\vspace{-.5cm}
\label{fig:R_and_Rout_over_Rin}
\end{figure}

Figure~\ref{fig:R_and_Rout_over_Rin} shows both the achievable outer rate $R_\mathrm{out}$ and the corresponding overall rate $R$ as a function of the inner rate $R_\mathrm{in}$ for $c=2$. The figure also illustrates the asymptotic behavior ($K \to \infty$), which follows from the proof of Corollary~\ref{cor:achievable_rate_large_k}: For $K \to \infty$, the achievable outer rate, given by the probability $\Pr\left(\tfrac{1}{K} \sum_{i=1}^K C_{d_i}(\Rix)>\Rin \right)$, approaches $1$ for $\Rin < \Exp(C_{d_i}(\Rix))$ and $0$ for larger $\Rin$.
As we have $R = \Rin \Rout (1-\nicefrac{\beta}{\Rix})$, for $K \to \infty$, we get a linear increase of the overall rate up to $\Rin = \Exp(C_{d_i}(\Rix))$ and $R = 0$ for larger $\Rin$.
As expected, the curves for approach the asymptotic behavior for growing $K$. An important interpretation of the ``$R$ over $\Rin$'' plot is that the positions of maxima of the curves give a design criterion on how to choose $\Rin$ to maximize $R$.

\bibliographystyle{IEEEtran}
\bibliography{ConcatenatedCodesforDNAStorage}

\end{document}